# Multivariate Ratio Estimation With Known Population Proportion Of Two Auxiliary Characters For Finite Population

*Rajesh Singh, *Sachin Malik, **A. A. Adewara, ***Florentin Smarandache

*Department of Statistics, Banaras Hindu University,Varanasi-221005, India
** Department of Statistics, University of Ilorin, Ilorin, Kwara State, Nigeria
*** Chair of Department of Mathematics, University of New Mexico, Gallup, USA

**Abstract**

In the present study, we propose estimators based on geometric and harmonic mean for estimating population mean using information on two auxiliary attributes in simple random sampling. We have shown that, when we have multi-auxiliary attributes, estimators based on geometric mean and harmonic mean are less biased than Olkin (1958), Naik and Gupta (1996) and Singh (1967) type- estimator under certain conditions. However, the MSE of Olkin( 1958) estimator and geometric and harmonic estimators are same up to the first order of approximation.

## 1. Introduction

Prior knowledge about population mean along with coefficient of variation of the population of an auxiliary variable is known to be very useful particularly when the ratio, product and regression estimators are used for estimation of population mean of a variable of interest. There exist situations when information is available in the form of the attribute $\phi$ which is highly correlated with y. For example y may be the use of drugs and $\phi$ may be gender. Using the information of point biserial correlation between the study variable and the auxiliary attribute Naik and Gupta (1996), Shabbir and Gupta (2006), Ab-Alfatah *et al.* (2010) and Singh *et al.* (2007, 2008) have suggested improved estimators for estimating unknown population mean $\overline{Y}$.

Using information on multi-auxiliary variables positively correlated with the study variable, Olkin (1958) suggested a multivariate ratio estimator of the population mean $\bar{Y}$. In this paper, we have suggested some estimators using information on multi-auxiliary attributes. Following Olkin (1958), we define an estimator as

$$\bar{y}_{ap} = \sum_{i=1}^{k} w_i r_i P_i \qquad (1.1)$$

where (i) $w_i$'s are weights such that $\sum_{i=1}^{k} w_i = 1$ (ii) $P_i$'s are the proportion of the auxiliary attribute and assumed to be known and (iii) $r_i = \dfrac{\bar{y}}{p_i}$, $\bar{y}$ is the sample mean of the study variable Y and $p_i$ is the proportion of auxiliary attributes $P_i$ based on a simple random sample of size n drawn without replacement from a population of size N.

Following Naik and Gupta (1996) and Singh *et al.* (2007), we propose another estimator $t_s$ as

$$t_s = \prod_{i=1}^{k} r_i P_i \qquad (1.2)$$

Two alternative estimators based on geometric mean and harmonic mean are suggested as

$$\bar{y}_{gp} = \prod_{i=1}^{k} (r_i P_i)^{w_i} \qquad (1.3)$$

and

$$\bar{y}_{hp} = \left( \sum_{i=1}^{k} \frac{w_i}{r_i P_i} \right)^{-1} \qquad (1.4)$$

such that $\sum_{i=1}^{k} w_i = 1$

These estimators are based on the assumptions that the auxiliary attributes are positively correlated with Y. Let $\rho_{\phi ij}$ (i=1,2,…k; j=1,2,…k ) be the phi correlation coefficient between $P_i$ and $P_j$ and $\rho_{0i}$ be the correlation coefficient between Y and $P_i$.

$$S_{\phi_{ij}}^2 = \frac{1}{N-1}\sum_{i=1}^{N}(\phi_{ji}-P_i)^2, \quad S_o^2 = \frac{1}{N-1}\sum_{i=1}^{N}(Y_i-\overline{Y})^2 \text{ and } C_0^2 = \frac{S_0^2}{\overline{Y}^2}, \quad C_i^2 = \frac{S_{\phi_{ij}}^2}{P_i^2}$$

In the same way $C_{0i}$ and $C_{ij}$ are defined.

Further, let $\underset{\sim}{w}' = (w_1, w_2, ..., w_k)$ and $C = [C_{ij}]_{p \times p}$ (i = 1,2,...k; j = 1,2,..., k)

## 2. BIAS AND MSE OF THE ESTIMATORS

To obtain the bias and MSE's of the estimators, up to first order of approximation, let

$$e_0 = \frac{\overline{y}-\overline{Y}}{\overline{Y}} \text{ and } e_i = \frac{p_i - P_i}{P_i}$$

such that $E(e_i) = 0$ (i = 0,1,2,...,k)..

Expressing equation (1.1) in terms of e's, we have

$$\overline{y}_{ap} = \sum_{i=1}^{k} w_i \overline{Y}(1+e_0)(1+e_i)^{-1}$$

$$= \overline{Y}\sum_{i=1}^{k} w_i [1 + e_0 - e_i + e_i^2 - e_0 e_i + e_0 e_i^2 - e_i^3] \quad (2.1)$$

Subtracting $\overline{Y}$ from both the sides of equation (2.1) and then taking expectation of both sides, we get the bias of the estimator $\overline{y}_{ap}$ up to the first order of approximation as

$$B(\overline{y}_{ap}) = f\overline{Y}\left[\sum_{i=1}^{k} w_i C_i^2 - \sum_{i=1}^{k} w_i C_{0i}\right] \quad (2.2)$$

Subtracting $\overline{Y}$ from both the sides of equation (2.1) squaring and then taking expectation of both sides, we get the bias of the estimator $\overline{y}_{ap}$ up to the first order of approximation as

$$MSE(\overline{y}_{ap}) = f\overline{Y}^2\left[C_0^2 + \sum_{i=1}^{k} w_i^2 C_i^2 - 2\sum_{i=1}^{k} w_i C_0 C_i + 2\sum\sum w_i w_j C_{ij}\right] \quad (2.3)$$

To obtain the bias and MSE of $\bar{y}_{gp}$ to the first order of approximation, we express equation (1.3) in term of e's, as

$$\bar{y}_{gp} = \prod_{i=1}^{k}\left[\bar{Y}(1+e_0)(1+e_i)^{-1}\right]^{w_i}$$

$$= \prod_{i=1}^{k}\bar{Y}\left[1+e_0 - w_i(e_i + e_0 e_i) + \frac{w_i(w_i+1)}{2}(e_i^2 + e_0 e_i^2)\right] \quad (2.4)$$

Subtracting $\bar{Y}$ from both sides of equation (2.4) and then taking expectation of both sides, we get the bias of the estimator $t_{gp}$ up to the first order of approximation, as

$$B(\bar{y}_{gp}) = f\bar{Y}\left[\sum \frac{w_i(w_i+1)C_i^2}{2} + \sum\sum w_i w_j C_{ij} - \sum w_i C_{0i}\right] \quad (2.5)$$

Subtracting $\bar{Y}$ from both the sides of equation (2.4) squaring and then taking expectation of both sides, we get the bias of the estimator $\bar{y}_{gp}$ up to the first order of approximation as

$$MSE(\bar{y}_{gp}) = f\bar{Y}^2\left[C_0^2 + \sum_{i=1}^{k}w_I^2 C_i^2 - 2\sum_{i=1}^{k}w_i C_0 C_i + 2\sum\sum w_i w_j C_{ij}\right] \quad (2.6)$$

Now expressing equation (1.4) in terms of e's, we have

$$\bar{y}_{hp} = \sum_{i=1}^{k} w_i \bar{Y}(1+e_0)(1+e_i)^{-1}$$

$$= \sum_i \bar{Y}(1+e_0)\left[1 - w_i e_i + w_i e_i^2 - w_i e_i^3\right]$$

$$= \bar{Y}\left[1 + e_0 - \sum_i w_i(e_i + e_0 e_i) + \left(\sum_i w_i e_i\right)^2\right] \quad (2.7)$$

Subtracting $\bar{Y}$ from both sides of equation (2.7) and then taking expectation of both sides, we get the bias of the estimator $\bar{y}_{hp}$ up to the first order of approximation will be

$$B(\bar{y}_{hp}) = f\bar{Y}\left[\sum_i w_i C_i^2 - \sum_i w_i C_{0i} + 2\sum\sum w_i w_j C_{ij}\right] \quad (2.8)$$

Subtracting $\overline{Y}$ from both the sides of equation (2.7) squaring and then taking expectation of both sides, we get the bias of the estimator $\overline{y}_{hp}$ up to the first order of approximation as

$$\mathrm{MSE}(\overline{y}_{hp}) = f\overline{Y}^2 \left[ C_0^2 + \sum_{i=1}^{k} w_i^2 C_i^2 - 2\sum_{i=1}^{k} w_i C_{0i} + 2\sum\sum w_i w_j C_{ij} \right] \tag{2.9}$$

We see that MSE's of these estimators are same and the biases are different. In general,

$$\mathrm{MSE}(\overline{y}_{gp}) = \mathrm{MSE}(\overline{y}_{hp}) = \mathrm{MSE}(\overline{y}_{ap}). \tag{2.10}$$

## 3. Comparison of biases

The biases may be either positive or negative. So, for comparison, we have compared the absolute biases of the estimates when these are more efficient than the sample mean. The bias of the estimator of geometric mean is smaller than that of arithmetic mean

$$\left| B(\overline{y}_{ap}) \right| > \left| B(\overline{y}_{gp}) \right| \tag{3.1}$$

Squaring and simplifying (3.1), we observe that

$$\left[ \frac{1}{2}\sum_{i=1}^{k} w_i^2 C_i^2 - 2\sum_{i=1}^{k} w_i C_{0i} + 2\sum\sum w_i w_j C_{ij} + \frac{3}{2}\sum_{i=1}^{k} w_i C_i^2 \right] \times$$

$$\left[ \frac{1}{2}\sum_{l=1}^{k} w_i C_i^2 - \frac{1}{2}\sum_{i=1}^{k} w_i^2 C_i^2 - \sum\sum w_i w_j C_{ij} \right] > 0 \tag{3.2}$$

Thus above inequality is true when both the factors are either positive or negative. The first factor of (3.2)

$$\left[ \frac{1}{2}\sum_{i=1}^{k} w_i^2 C_i^2 - 2\sum_{i=1}^{k} w_i C_{0i} + 2\sum\sum w_i w_j C_{ij} + \frac{3}{2}\sum_{i=1}^{k} w_i C_i^2 \right]$$

is positive, when

$$\frac{\sum_{i=1}^{k} w_i^2 C_i^2}{\underset{\sim}{w}' C \underset{\sim}{w}} > \frac{1}{3} \tag{3.3}$$

In the same way, it can be shown that the second factor of (3.2) is also positive when

$$\frac{\sum_{i=1}^{k} w_i^2 C_i^2}{w' C w} > 1 \qquad (3.4)$$

When both the factors of (3.2) is negative, the sign of inequalities of (3.3) and (3.4) reversed.

Also comparing the square of the biases of geometric and harmonic estimator, we find that geometric estimator is more biased than harmonic estimator.

Hence we may conclude that under the situations where arithmetic, geometric and harmonic estimator are more efficient than sample mean and the relation (3.4) or

$$\frac{\sum_{i=1}^{k} w_i^2 C_i^2}{w' C w} < \frac{1}{3}$$

is satisfied, the biases of the estimates satisfy the relation

$$\left| B(\bar{y}_{ap}) \right| > \left| B(\bar{y}_{gp}) \right| > \left| B(\bar{y}_{hp}) \right|$$

Usually the weights of $w_i$'s are so chosen so as to minimize the MSE of an estimator subject to the condition

$$\sum_{i=1}^{k} w_i = 1.$$

## 4. Empirical Study

Data : (Source: Singh and Chaudhary (1986), P. 177).

The population consists of 34 wheat farms in 34 villages in certain region of India. The variables are defined as:

y = area under wheat crop (in acres) during 1974.

$p_1$ = proportion of farms under wheat crop which have more than 500 acres land during 1971.

and

$p_2$ = proportion of farms under wheat crop which have more than 100 acres land during 1973.

For this data, we have

N=34, $\bar{Y}$=199.4, $P_1$=0.6765, $P_2$=0.7353, $S_y^2$=22564.6, $S_{\phi_1}^2$=0.225490, $S_{\phi_2}^2$=0.200535, $\rho_{pb_1}$=0.599, $\rho_{pb_2}$=0.559, $\rho_\phi$=0.725.

Biases and MSE's of different estimators under comparison, based on the above data are given in Table 4.1.

**TABLE 4.1 : Bias and MSE of different estimators**

| Estimators | Auxiliary attributes | Bias | MSE |
|---|---|---|---|
| $\bar{y}$ | none | 0 | 1569.795 |
| Ratio $\bar{y}\left(\dfrac{P_1}{p_1}\right)$ | $P_1$ | 2.4767 | 1197.675 |
| Ratio $\bar{y}\left(\dfrac{P_2}{p_2}\right)$ | $P_2$ | 1.6107 | 1194.172 |
| Olkin ($\bar{y}_{ap}$) | $P_1$ and $P_2$ | 2.0415 | 1024.889 |
| Suggested $\bar{y}_{gp}$ | $P_1$ and $P_2$ | 1.6126 | 1024.889 |
| Suggested $\bar{y}_{hp}$ | $P_1$ and $P_2$ | 1.1838 | 1024.889 |
| $t_s = \bar{y}\left(\dfrac{P_1}{p_1}\right)\left(\dfrac{P_2}{p_2}\right)$ | $P_1$ and $P_2$ | 8.4498 | 2538.763 |

## 5. Conclusion

From Table 4.1 we observe that the MSE's of Olkin (1958) type estimator, estimator based on harmonic and geometric mean are same. Singh (1967) type estimator $t_s$ performs worse. However, the bias of the ratio-type estimator based on harmonic mean is least. Hence, we may conclude that when more than one auxiliary attributes are used for estimating the population parameter, it is better to use harmonic mean.

**References.**

Abd-Elfattah, A.M. El-Sherpieny, E.A. Mohamed, S.M. Abdou, O. F. (2010): Improvement in estimating the population mean in simple random sampling using information on auxiliary attribute. Appl. Math. and Compt. doi:10.1016/j.amc.2009.12.041

Naik, V.D and Gupta, P.C. (1996): A note on estimation of mean with known population proportion of an auxiliary character. Jour. Ind. Soc. Agri. Stat., 48(2), 151-158.

Olkin, I. (1958): Multivariate ratio estimation for finite population . Biometrica, 45. 154-165.

Shabbir, J. and Gupta, S. (2006): A new estimator of population mean in stratified sampling, Commun. Stat. Theo. Meth. 35: 1201–1209

Singh, D. and Chaudhary, F. S. (1986) : Theory and Analysis of Sample Survey Designs John Wiley and Sons, NewYork.

Singh, R., Cauhan, P., Sawan, N. and Smarandache,F. (2007): Auxiliary information and a priori values in construction of improved estimators, Renaissance High press, USA.

Singh, R., Chauhan, P., Sawan, N. and Smarandache,F. (2008): Ratio estimators in simple random sampling using information on auxiliary attribute. Pak. J. Stat. Oper. Res.,4,1,47-53.

Singh, M.P. (1967): Multivariate product method of estimation for finite populations. J. Indian Soc. Agri. Statist., 19, 1-10.